# Efficient integration of self-assembled organic monolayer tunnel barriers in large area pinhole-free magnetic tunnel junctions

*Maryam S. Dehaghani [a, b], Sophie Guézo Aguinet [a], Arnaud Le Pottier [a], Soraya Ababou-Girard [a], Rozenn Bernard [a, c], Sylvain Tricot [a], Philippe Schieffer [a], Bruno Lépine [a], Francine Solal [a], Pascal Turban [a, *]*

[a] Université de Rennes, CNRS, IPR(Institut de Physique de Rennes) - UMR 6251, 35000 Rennes, France

[b] Université de Toulouse, LAAS-CNRS, 31400 Toulouse, France

[c] Université de Rennes, INSA de Rennes, CNRS, Institut FOTON - UMR 6082, 35000 Rennes, France

**ABSTRACT:** Magneto-transport properties in hybrid magnetic tunnel junctions (MTJs) integrating self-assembled monolayers (SAMs) as tunnel barriers are critically influenced by spinterface effects, which arise from the electronic properties at ferromagnet (FM)/SAM interfaces. Understanding the mechanisms governing spinterface formation in well-controlled model systems is essential for the rational design of efficient molecular spintronic devices. However, the fabrication of FM/SAM/FM systems remains a significant challenge due to the difficulty in preventing electrical shorts through the SAM tunnel barrier during top FM electrode deposition. In this study, we address these challenges by developing model hybrid MTJs incorporating alkanethiol SAM tunnel barriers grafted under ultra-high vacuum conditions onto single-crystalline Fe(001) bottom electrodes. A soft-landing deposition method is used for the deposition of a top Co FM electrode. The deposition process and the electronic properties of the FM/SAM interfaces are first studied by spatially integrated X-ray photoelectron spectroscopy. Furthermore, ballistic electron emission microscopy (BEEM) and spectroscopy are used to investigate the lateral homogeneity of the organic barrier. Optimal soft-landing deposition conditions allows the preparation of homogeneous Co/SAM interfaces with no evidence of metal diffusion through the SAM at the nanoscale. These observations are further confirmed at the micron-scale by the high-yield patterning of large area (5×5µm$^2$) MTJs presenting fingerprints of electron tunneling through the SAM. These findings provide critical insights into the fabrication and optimization of molecular spintronic devices, paving the way for advancements in hybrid MTJ technology.

1. **INTRODUCTION**

Since the pioneering experimental works by Dediu et al. [1] and Xiong et al. [2], molecular spintronics has attracted increasing scientific interest [3,4]. The integration of molecular materials in spintronic devices was initially motivated by the longer spin lifetime in organic layers due to the combined low-spin-orbit coupling and hyperfine interactions in these systems [5]. Magnetoresistance properties of hybrid spin-valves or magnetic tunnel junctions were further observed to be strongly correlated with the electronic properties of the hybrid organic/inorganic interfaces of the devices giving birth to the so-called spinterface concept [6,7]. Spinterface effects are intimately depending on the fine interface electronic coupling between molecular frontier orbitals and electronic spin-polarized states close to the Fermi level of the ferromagnetic electrodes. Experimentally, such spin-dependent hybridized interface states have been shown to govern magnetotransport properties in hybrid magnetic tunnel junctions (MTJs), leading for example to enhanced magnetoresistance amplitude or even magnetoresistance sign inversion[7–9]. Spinterface effects are thus promising to tailor the properties of spin-injection at hybrid ferromagnet/molecule interfaces towards efficient molecular spintronics devices. However, taking advantage of spinterface effects in hybrid magnetic tunnel junctions integrating a self-assembled organic monolayer (SAM) is still hampered by major experimental constrains. First, the oldest and most crucial challenging aspect of hybrid MTJs fabrication is to avoid the formation of electrical shorts through the molecular layer during the top ferromagnetic electrode deposition. Metal diffusion through the SAM is indeed commonly observed when depositing a metallic thin film on a SAM, leading to pinhole formation redhibitory for spin-dependent tunneling. This issue of a reliable metal top-contact to electrically address SAM is very general, and various alternative technical solutions to electrically contact SAM were proposed in the general framework of molecular electronics [10], such as metal layer transfer printing [11,12], cross wires junctions [13,14], spin-coating of a conductive polymer contact [15,16], nanopore junctions [17,18], use of high surface energy metals [19–22]. However, these methods are not always compatible with large scale integration and/or cannot be used with highly reactive metallic electrodes such as ferromagnetic transition metals. So far, the most successful strategy to study magnetotransport properties of functional hybrid MTJs with a SAM barrier is based on specific patterning methods leading to sufficiently small junction areas with non-zero probability to be pinhole-free. For instance, nanosphere lithography has been employed to reliably fabricate submicron Co/CoPc/Co MTJs (diameter 300 nm) using $SiO_2$ nanobead masks to pattern nanopillars[23]. Alternatively, a nanoindentation lithography technique was successfully used by Tatay et al.[24] to prepare Co/SAM/$La_{2/3}Sr_{1/3}MnO_3$ MTJs with diameter in the 10-50nm range. Even for this nanometric junction size, typically only 25% of the prepared junctions presented non-linear I(V) curves typical for electron tunneling. Larger molecular MTJs with junction diameter in the 400-800nm range could also be patterned using laser lithography while achieving a 30% ratio of non-shortened junctions [25]. At this point, the development of an adapted deposition process enabling preparation of large area pinhole-free organic magnetic tunnel junctions with a high yield and well-controlled model spinterfaces is still a technological bottleneck that has to be addressed.

In the present work, we present an experimental study of model hybrid Co/SAM/Fe(001) magnetic tunnel junctions integrating a 1-hexadecanethiol SAM tunnel barrier. The single-crystalline bottom Fe(001) electrode, commonly used in inorganic MTJs [26,27] and with well-known electronic properties [28,29] is an ideal platform for future investigation of spinterface effects. Furthermore, SAMs grafted on single-crystalline Fe(001) electrode present well-defined geometry with respect to the ferromagnet surface lattice [30]. Besides, in these systems, the degree of electronic coupling between molecules and ferromagnet at this bottom interface can also conceptually be modulated by changing the nature of the anchor group while the degree of electronic delocalization through the organic backbone can be varied from an insulating organic tunnel barrier (for saturated chains) to a more conductive SAM (for unsaturated chains). Finally, our MTJ growth process is fully handled under ultra-high vacuum environment, avoiding any surface/interface oxidation of the highly reactive ferromagnet electrodes leading to clean and well-defined interfaces. We thus believe that the developed methodology is promising for achieving model hybrid MTJs suitable to tackle spinterface effects. We have specifically addressed the issue of pinhole formation during molecular beam epitaxy (MBE) deposition of the ferromagnetic Co top electrode on the SAM tunnel barrier by using a soft-landing method which was previously demonstrated to reduce metal penetration through thick organic layers [31]. The cobalt soft-landing deposition was first validated on a reference SAM grafted on a GaAs(001) substrate. This first sample allowed us to develop a multiscale analysis based on spatially averaged X-ray photoelectron spectroscopy (XPS) and nanometer-scale ballistic electron emission microscopy (BEEM) experiments, and to compare Co/SAM/GaAs(001) contacts prepared by room-temperature MBE deposition and by soft-landing method. While a dramatic Co diffusion through the SAM is observed at room-temperature, the proposed soft-landing method results in homogenous Co/SAM interfaces with a preserved organic tunnel barrier over extended areas. Finally, the soft landing method is further combined with shadow mask deposition under ultra-high vacuum to pattern hybrid Co/SAM/Fe(001) MTJs grown on MgO(001). 44% of the prepared large-area (25µm$^2$) tunnel junctions present non-linear I(V) curves typical of electron tunneling through a preserved SAM confirming the validity of this method for efficient SAM-based spintronic device preparation.

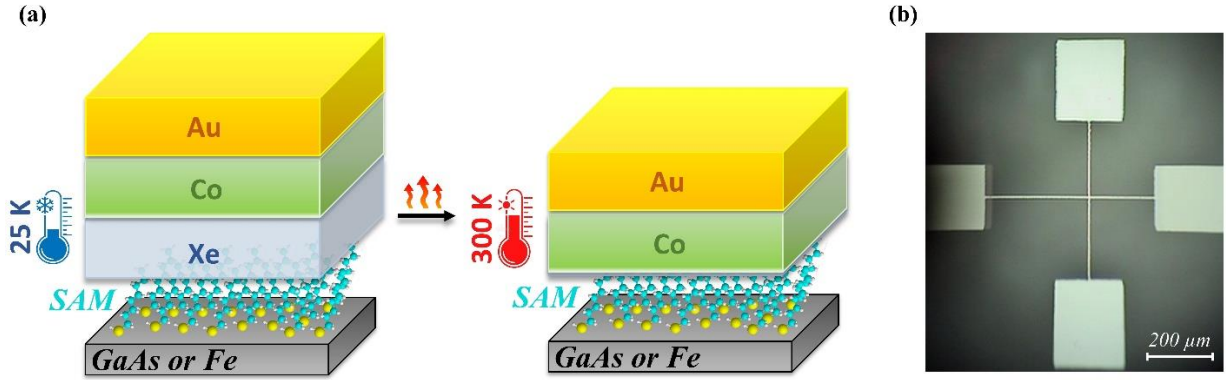

**Figure 1.** (a) Schematic sketch of the soft-landing technique for Au/Co top contact deposition on a SAM tunnel barrier, (b) Optical micrograph top view of a 5×5μm² Au/Co/SAM/Fe(001)/MgO(001) magnetic tunnel junction in the crossbar geometry fully grown under ultra-high vacuum by combined soft-landing and shadow mask deposition.

## 2. RESULTS AND DISCUSSION

**Soft-landing deposition of Co ferromagnetic electrodes.** We have grown cobalt ferromagnetic top contact by MBE on dense 1-hexadecanethiol (HDT, $C_{16}H_{33}$-SH) SAM grafted on a Fe(001) bottom electrode grown by molecular beam epitaxy on MgO(001) substrate for the preparation of Co/HDT/Fe(001) magnetic tunnel junctions. Alternatively, identical Co top electrodes were also deposited on HDT SAM grafted on GaAs(001) to allow efficient BEEM and XPS characterizations of the Co/HDT contact homogeneity. The Co layers were evaporated from a Knudsen cell, either on sample maintained at room temperature (RT), or using a soft-landing deposition process [32]. In this case, after SAM grafting, sample was cooled down at 25K prior exposure to 900 Langmuir of Xe, forming a 100nm-thick protective Xe buffer layer on the SAM before Co deposition. Finally, sample was slowly warmed up to RT to gently desorb the Xe layer (Figure 1a). This soft-landing deposition process was combined with in situ shadow mask deposition to pattern magnetic tunnel junctions in the crossbar geometry integrating a HDT SAM tunnel barrier (Figure 1b). Finally, a 4nm thick Au caping layer was finally deposited on top of the cobalt electrode to prevent oxidation. Reference Au/Co/GaAs(001) Schottky contacts without SAM were also grown using the same protocol at RT or using soft-landing for comparison with samples integrating the HDT tunnel barrier.

**XPS characterization.** X-ray photoemission spectroscopy (XPS) was first used as a spatially integrated diagnosis to evaluate the Co top-electrode penetration through a HDT monolayer grafted on GaAs(001). Figure 2a presents the Ga 3d core level recorded after RT deposition of a 1.2nm thick Co layer on a bare GaAs(001) surface. For this RT-grown reference Schottky contact the Ga 3d core level presents two well separated components. The main peak at 19.45 eV of binding energy is associated to Ga atoms in bulk GaAs as confirmed on a reference spectra on the free GaAs(001) surface. The second component, shifted by 1eV towards lower binding energy is associated to Ga atoms in a metallic environment due to the formation of a CoGa alloy at the Co/GaAs interface [33]. A very similar Ga 3d line shape is obtained after RT Co deposition on the HDT SAM grafted on GaAs(001) as shown in Figure 2b. Specifically, the CoGa reaction component is also observed after Co deposition on the HDT SAM grafted on GaAs(001), with similar intensity as on the reference Co/GaAs(001) Schottky contact. This observation unambiguously demonstrates that Co diffuses significantly through the molecular barrier during the room temperature deposition process, leading to extended contact areas between Co and GaAs where alloying occurs. On the other hand, as expected for soft-landing deposition [32], the absorption of the cobalt adatoms on a solid Xenon buffer layer promotes the formation of an abrupt Co/GaAs interface, as evidenced by XPS. The metallic reaction component associated to the CoGa alloy is indeed completely absent in both the reference Co/GaAs(001) Schottky contact (figure 2c) and Co/HDT/GaAs(001) sample (figure 2d) prepared by soft-landing, which demonstrates an effective approach for achieving a well-controlled interface. At this point, it should be emphasized that the absence of CoGa alloy formation in the Co/HDT/GaAs(001) sample prepared by soft-landing might not provide conclusive information on the lateral homogeneity of the Co/SAM top-contact, nor on the absence of pinholes formation using this deposition process. Specifically, one cannot exclude on the basis these XPS observations that metallic shorts are still formed through the HDT SAM during soft-landing deposition but with a now local abrupt Co/GaAs interface in the pinhole regions. Complementary characterizations are thus required to study the lateral homogeneity of the Co/HDT contact at the nanoscale.

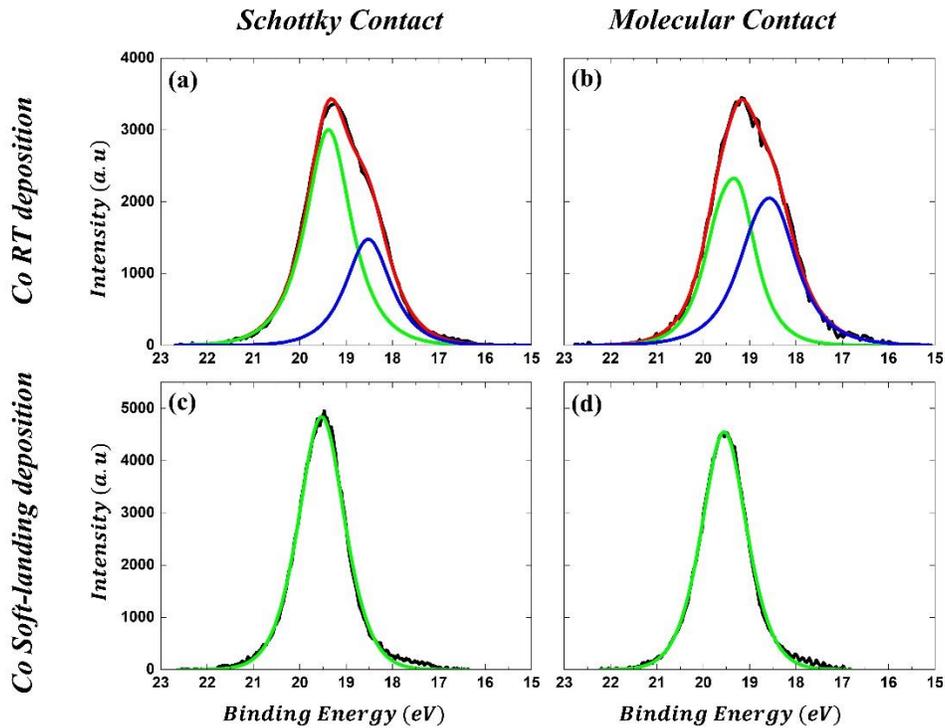

**Figure 2.** XPS study of Ga 3d evolution after Co deposition (12 Å): (a, b) Co/GaAs(001) and Co/SAM/GaAs(001) at room temperature, and (c, d) Co/GaAs(001) and Co/SAM/GaAs(001) employing soft-landing method. Black line correspond to the raw experimental data, colored lines to a fit of the experimental data. The bulk component of the GaAs substrate (shown in green) appears at 19.5 eV, while the metallic reaction component (shown in blue) is observed at 18.5 eV.

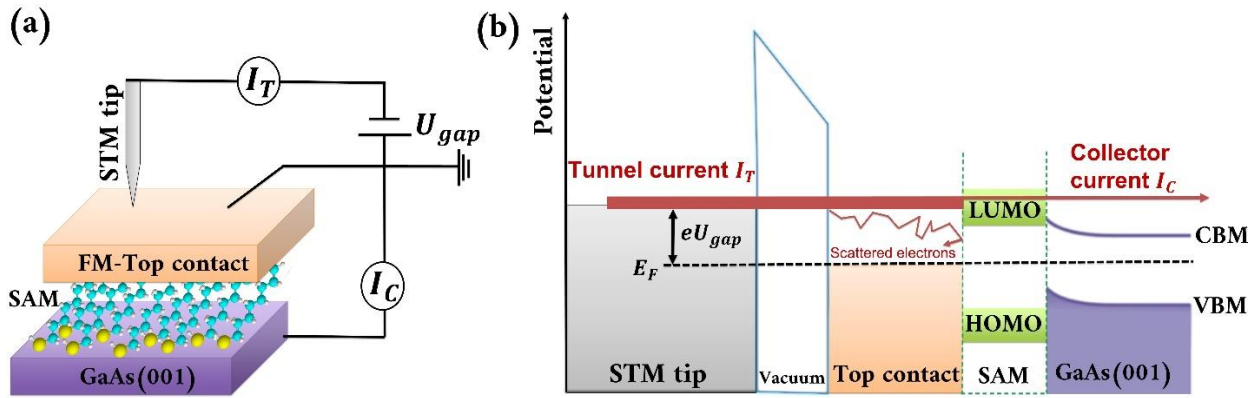

**Figure 3.** Principle of a BEEM experiment on a Metal/SAM/GaAs system, (a) A schematic of BEEM experimental setup. (b) so-called ballistic electrons can transport through the SAM LUMO states if their energy is large enough and be collected in the GaAs semiconductor substrate.

**Nanoscale BEEM characterization.** Ballistic Electron Emission Microscopy (BEEM) was conducted to assess the spatial homogeneity of the HDT organic tunnel barrier after Co top contact deposition. In a BEEM experiment on a metal/SAM/n-GaAs sample (Figure 3), a hot-electron current (noted $I_T$) is locally injected from the STM tip at the metal surface with an energy defined by the applied tip/sample tunnel bias $U_{gap}$. After transport through the metal layer, the hot-electrons beam can further propagate through the unoccupied molecular orbital states of the SAM, provided the electron injection energy overcomes the barrier height at the metal/SAM interface, and finally enter in the conduction band of the GaAs. This collector current ($I_c$) can be measured via an ohmic contact on the GaAs substrate backside. The obtained collector current image is thus a map of sample electronic transmission at a specific electron energy making BEEM a powerful tool allowing nanoscale imaging of transport inhomogeneities in inorganic [34] or SAM [35] tunnel barriers. Furthermore, the initial energy threshold for the hot-electron current onset in the BEEM spectroscopy curves $I_c(U_{gap})$ gives access to the local barrier height while further inflection of the BEEM spectra allows identification of eventual new transport channel opening at higher energy.

We first compared the electron transmission through a reference Au(4nm)/Co(3nm)/GaAs(001) Schottky contacts and a Au(4nm)/Co(3nm)/SAM/GaAs(001) hybrid tunnel contact for which Co deposition was performed at RT. Figure 4 presents STM and simultaneously measured BEEM images on both Schottky and SAM tunnel contact. On the Schottky contact, surface morphology (Figure 4a) is typical of an epitaxial Au/Co bilayer grown on GaAs(001) with the presence of atomic terraces and step edges running along the crystal high symmetry direction. A square atomic lattice is observed on the Au terraces typical of a Au(001) overlayer grown on epitaxial Co/GaAs(001). The associated BEEM image (Figure 4b) presents, aside measurement noise, a homogeneous current distribution over the field of observation. The surface morphology of the metal bilayer grown on the HDT SAM is drastically different (Figure 4d). A granular morphology, with lateral grain size of 5-10nm and peak to peak roughness of 1nm, typical of a polycrystalline film, is observed. The corresponding BEEM image (Figure 4e) presents a hot-electron transmission map with low-amplitude current modulations at the nanoscale which are clearly correlated with the film morphology. Electron inelastic scattering processes occurring during propagation through the Au/Co bilayer leads to an exponential attenuation of the hot-electron beam with metal thickness [36]. The local metal roughness thus modulates the electron transmission resulting in the observed BEEM current variations. Aside from these weak modulations purely induced by local metal thickness variations, the RT-grown Au/Co/SAM/GaAs contact presents thus a homogeneous electron transmission. No local high/low electron transmission contrasts can be found at the interface, excluding the coexistence of localized metal shorts surrounded by preserved SAM tunnel barrier regions which would result in large-amplitude current modulations, easily imaged by BEEM [35]. Similar homogeneous BEEM images could be reproducibly observed on various measurement locations on a single junction, as well as on other junctions patterned over a 1cm² sample.

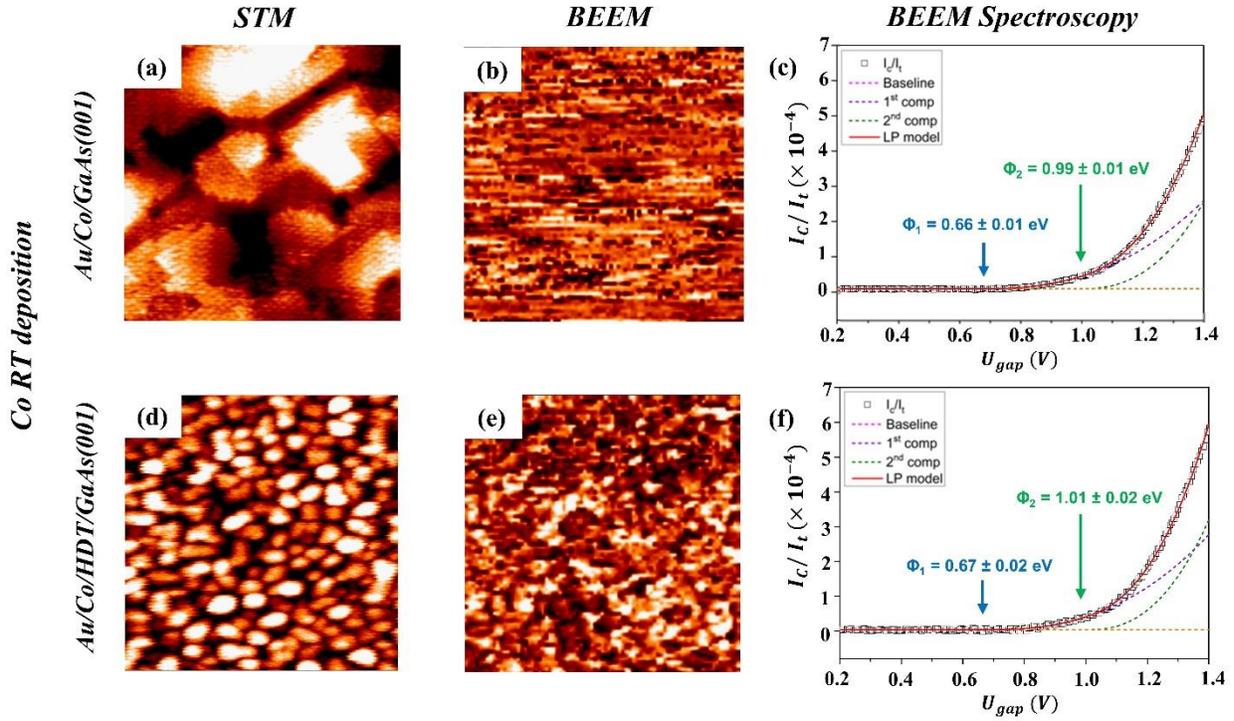

**Figure 4.** Room temperature deposition (a) 25 × 25 $nm^2$ STM image ($U_{gap}$=0.038 V, $I_T$=10 nA, Z scale 0 to 522 pm) and (b) simultaneously recorded BEEM image ($U_{gap}$=1.72 V, $I_T$=11nA, current scale 0 to 9 pA) on a Au(4nm)/Co(3nm)/GaAs(001) sample. (d) 75 × 75 $nm^2$ STM image ($U_{gap}$=1.75 V, $I_T$=15.3 nA, Z scale 0 to 0.8 nm) and (e) simultaneously recorded BEEM images ($U_{gap}$=1.75 V, $I_T$=15.3 nA, current scale 0 to 40 pA) of Au(4nm)/Co(3nm)/SAM/GaAs(001). Local BEEM spectroscopy curves (average of 200 measurement points over a typical 75 × 75$nm^2$ surface) measured on (c) Au/Co/GaAs(001) and (f) Au/Co/SAM/GaAs(001) samples.

To go one step further in the analysis, BEEM spectroscopy curves measured on both RT-grown Schottky and SAM tunnel contacts are compared in Figures 4c and f. 400 individual spectra were averaged to improve the signal-to-noise ratio. The experimental data were fitted with the commonly used Ludeke-Prietsch (LP) 5/2 power-law [37]:

$$\frac{I_C}{I_T} = a_0 + a_1(E - e\Phi_1)^{5/2} + a_2(E - e\Phi_2)^{5/2} . \qquad (1)$$

For the Schottky contact, the first energy threshold $\phi_1$=0.66eV is nicely matching the Schottky barrier height measured on a RT grown Co/n-GaAs interface[32]. $\phi_1$ thus corresponds to the energy difference between the Co Fermi level and GaAs conduction band minimum located at the Γ point of GaAs Brillouin zone. The second threshold $\phi_2$=0.99eV is associated to the opening of a new transport channel by injection of hot-electron in the L-valley of GaAs conduction band, which is located in the GaAs band structure 0.33eV above the Γ-valley [38], in excellent agreement with our BEEM data. BEEM spectroscopy curve measured on the RT-grown Au/Co/SAM/GaAs contact presents striking similarities with the reference Schottky contact. First, the overall magnitudes of the electron transmission are almost identical in both samples. There is no additional attenuation of the BEEM current as expected after the insertion of the molecular tunnel barrier at the Co/GaAs interface. The energy thresholds deduced from LP fit are also identical (within measurement uncertainties) further confirming identical transport mechanisms for both samples. From these BEEM observations at the nanoscale, we deduce that room-temperature Co deposition on the HDT-SAM results in extended diffusion through the organic monolayer. This metal diffusion process is even more severe for Co deposition with our experimental parameters (see experimental details) compared to our previous BEEM observation on Au overlayers grown at RT on similar SAM for which preserved isolated molecular patches with nanometric lateral dimensions could be imaged locally by BEEM [35]. This is most likely due to the use of a high temperature effusion cell necessary for Co deposition generating significant radiative heating of sample surface. The resulting interface appears homogeneous without even any fingerprint of locally preserved molecular patches which is consistent with the previous XPS observations of large interfacial Co/GaAs intermixing through the SAM.

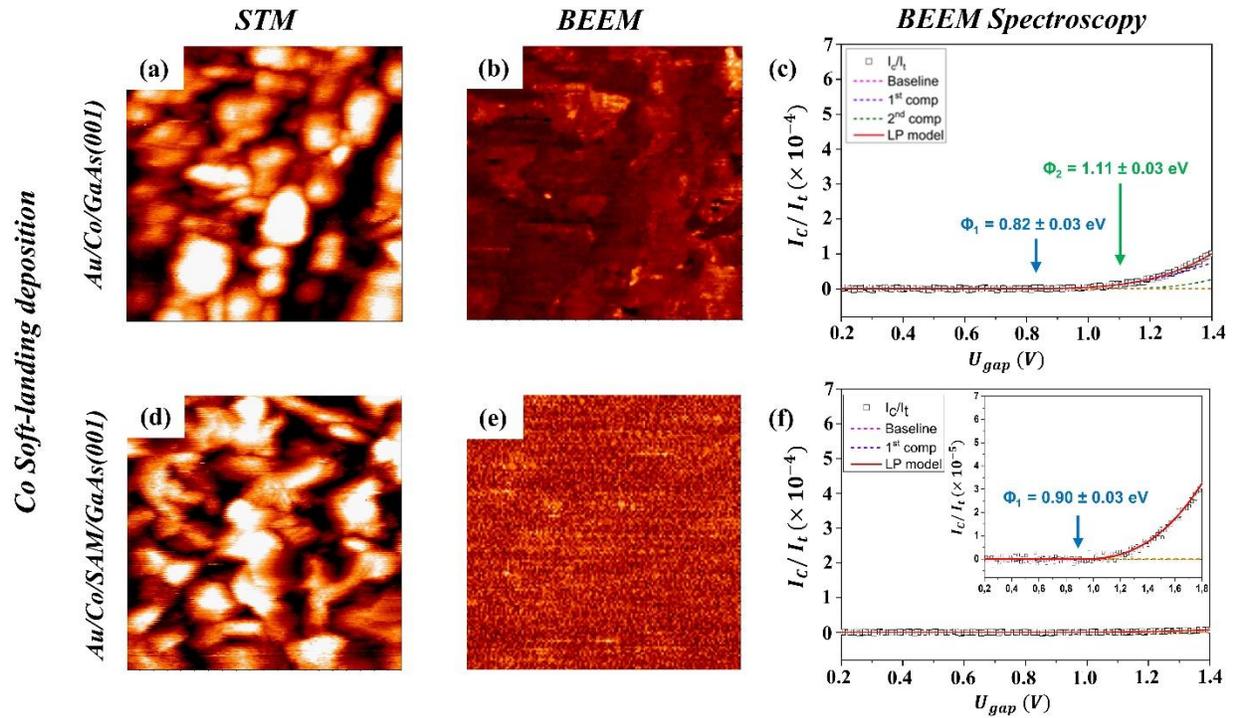

**Figure 5.** Soft-landing deposition (a) STM image and (b) simultaneously recorded BEEM image of Au(4nm)/Co(3nm)/GaAs(001) at 1.82 V, $I_T = 15 nA$ (color scale: 85 to 110 pA), $100 \times 100\ nm^2$. (d) and (e) STM and BEEM images of Au(4nm)/Co(3nm)/HDT/GaAs(001) at 1.81 V, $I_T = 10 nA$ (color scale: 16 to 20 pA), $100 \times 100\ nm^2$. Local BEEM spectroscopy curves (average of 400 measurement points over a typical $75 \times 75 nm^2$ surface) measured on (c) Au/Co/GaAs(001) sample at $I_T = 15.0\ A$, and (f) Au/Co/SAM/GaAs(001) sample at $I_T = 15.0\ A$, the insert provides a magnified view of the onset of the BEEM signal, highlighting the threshold behavior.

A similar comparative BEEM study was performed on a reference Au(4nm)/Co(3nm)/GaAs(001) Schottky contact and a Au(4nm)/Co(3nm)/SAM/GaAs(001) hybrid tunnel contact prepared this time by soft-landing (Figure 5). Metal deposition on the solid Xe buffer layer results for both samples in polycrystalline metal layers with typical lateral grain size between 8 and 15nm as observed by STM (Figures 5a and 5d) and a typical peak-to peak roughness of 4nm. The simultaneously collected BEEM current images for both samples are presented in Figure 5b and 5e. The electron transmission through the Au/Co/GaAs(001) Schottky contact presents moderate local variations correlated with the grain morphology. The BEEM image for the SAM-sample shows a homogeneous interface with a very low current level and no sign of pinholes. BEEM spectroscopy has been complementary used to explain the electron transport in the two samples prepared by soft-landing. Figure 5c presents a representative BEEM spectroscopy curve (average of 400 individual spectra) obtained on the reference Au/Co/GaAs Schottky. The Ludeke-Prietsch fit of the experimental data is also presented leading to a first threshold at $\phi_1$=0.82eV and a second component at $\phi_1$=1.11eV. The energy difference between the two thresholds is again in good agreement with the energy separation between the bottoms of the $\Gamma$ and L valleys of GaAs conduction band. The Schottky barrier height at the Co/GaAs(001) interface is thus determined to be 0.82eV for a sample prepared by soft-landing. The significant 0.16eV increase of Schottky barrier height compared to the room-temperature deposited Schottky contact is consistent with the fact that Co soft-landing deposition allows formation of an abrupt Co/GaAs interface, without formation of the CoGa interface alloy as observed by XPS (Figure 2c). Figure 5f shows a typical BEEM spectroscopy curve obtained on the Au(4nm)/Co(3nm)/SAM/GaAs(001) tunnel contact prepared by soft-landing. The overall magnitude of the BEEM signal on this sample has drastically decreased compared to the reference Schottky contact prepared by soft-landing. For instance, the BEEM current at 1.6eV is roughly 20 time larger on the Schottky contact than on the SAM sample. This additional strong attenuation process of the hot-electron beam compared to the Schottky contact is attributed to the presence of a preserved molecular layer at the interface by using soft-landing deposition. Further fingerprints of the preserved SAM tunnel barrier can be obtained from the Ludeke-Prietsch fit of the BEEM spectra. An accurate fit over the full energy window was achieved using only a single component with a threshold value $\phi_1$=0.90eV (see insert of Figure 5f). This value, significantly larger than the Schottky barrier height in the Au/Co/GaAs(001) reference Schottky contact prepared by soft-landing, is associated to the new energy position of the Γ-valley of GaAs conduction band after insertion of at the SAM at the Co/GaAs interface. Alkanethiols are indeed polar molecules with a typical dipole value estimated to be 2-3 Debye [39]. The Γ point is thus shifted 0.08eV higher in energy after insertion of the SAM due to the interface electrostatic dipole introduced by the molecules, as previously observed in the Au/SAM/GaAs(001) system [35]. Our previous investigation of Au/SAM/GaAs(001) samples prepared by soft-landing have shown BEEM spectra presenting two transport channels [35]. At low energy, a first transport channel by electron tunnelling trough the SAM into the Γ valley of GaAs was first observed. At higher energy, a clear second threshold was observed and attributed to the opening of a transport channel in the LUMO states of the SAM. Since with a cobalt top contact deposited on the molecular layer only one transport channel is observed, we assume that the LUMO states of the SAM are located lower in energy at the metal/molecule interface for the Co/SAM/GaAs system than for Au/SAM/GaAs. This idea is supported by the electronegativity values of the two considered metals which governs the Fermi level position at the metal/molecule interface [40]. Indeed, the significantly lower Pauling electronegativity for Co (1.88) than for Au (2.54) will favor a lower tunnel barrier at the Co/SAM interface than at the Au/SAM [41], and thus

LUMO states lower in energy. If the LUMO energy states lowers and overlaps the GaAs conduction band minimum, the low energy tunnel transport channel through the SAM will disappear, and a single transport channel is possible by direct electron propagation through the LUMO states to the Γ-valley of GaAs, as represented on the energy diagram of Figure 3b.

To conclude, BEEM imaging and spectroscopy performed at the nanoscale both point out a homogeneous and preserved SAM tunnel barrier after cobalt electrode deposition by soft-landing. Since similar observations could be reproduced at various measurement points of one junction as well as on different junctions, we expect that the SAM was efficiently protected on a large scale by the solid Xe layer during the cobalt electrode deposition by soft-landing. An efficient method to confirm this conclusion is to study the transport properties of magnetic tunnel junctions grown by soft-landing.

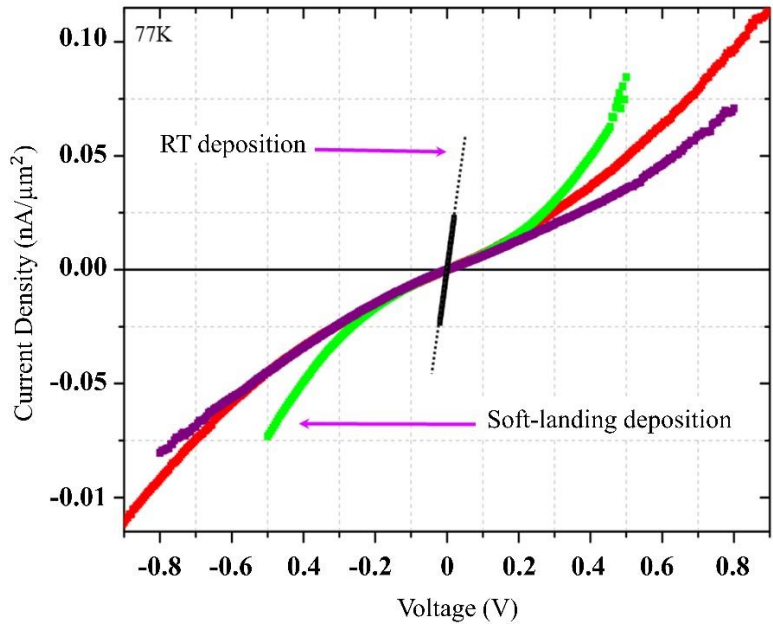

**Figure 6:** Comparison of typical J(V) curves measured at 77K on $5 \times 5 \mu m^2$ Au(4nm)/Co(15nm)/SAM/Fe(001)(30nm)/MgO(001) magnetic tunnel junctions for a Au/Co top electrode grown at room-temperature (black curve) or using the soft-landing process (colored curves).

**Electrical properties of Co/SAM/Fe(001) magnetic tunnel junctions prepared by soft-landing.** To further- assess the general applicability of the soft-landing approach beyond BEEM compatible systems, we extended our study to large-area tunnel junctions. We investigated whether soft-landing can be used to reliably form high-resistance pinhole-free SAM-based tunnel junctions. For this purpose, Au(4nm)/Co(15nm)/SAM/Fe(30nm)/MgO(001) hybrid magnetic tunnel junctions were patterned in the crossbar geometry using shadow mask deposition (see experimental section). Figure 6 presents typical current density versus applied bias J(V) curves measured at 77K on 5×5μm$^2$ junctions. For junctions prepared by soft-landing, high resistance values at low bias (varying between 4.9kΩ and 2.2MΩ) followed by nonlinear J(V) behavior at higher voltages are reproducibly observed. These J(V) traces, fingerprints of electron tunneling transport mechanism through the SAM, could be obtained with a 44% yield per sample. On the other hand, 100% of the junctions displayed ohmic behavior with a low resistance around zero bias (0.9kΩ) when the Co/Au top electrode was grown at RT. These transport results indicate that soft-landing efficiently prevents the formation of pinholes by metal diffusion through the SAM over large area hybrid magnetic tunnel junctions.

3. **CONCLUSIONS**

In summary, we have developed an efficient method to prepare model hybrid magnetic tunnel junctions integrating a self-assembled molecular monolayer as a tunnel barrier. Based on combined in situ structuration of junctions by shadow masks and soft-landing deposition of the ferromagnetic top electrode, this process is fully performed under ultrahigh vacuum environment and allows the preservation of well-controlled interfaces, a key point for future exploitation of spinterface effects in organic spintronic devices. A multiscale experimental study, combining spatially integrated X-ray photoelectron spectroscopy and transport measurement with nanoscale ballistic electron emission microscopy, demonstrates that soft-landing allows preparation of pinhole-free, highly homogeneous tunnel contacts over large (5×5μm$^2$) lateral dimensions. This methodology is applicable in principle to any organic monolayer and any top/bottom ferromagnetic electrode offering a large versatility for high yield organic spintronic device preparation.

## 4. METHODS

**Sample Preparation.** Ferromagnetic electrodes were evaporated by using Knudsen cells using cobalt (Co, ≥99.99 wt%) and iron (Fe, ≥99.99 wt%) shots purchased from MaTeck. Mechanically polished MgO(001) substrates were sourced from SurfaceNet. The organic molecule 1-hexadecanethiol (HDT, $C_{16}H_{33}$-SH, purity >95%) was supplied by Sigma-Aldrich and further purified by freezing/pumping cycles. Two different types of samples were fully prepared under ultra-high vacuum (UHV) environment (P<5*10$^{-10}$mbar) in this work. The Co metal top contact deposition process was first optimized on a HDT SAM grafted on GaAs(001) to allow BEEM characterization. A 1.5 μm-thick Si n-doped (4*10$^{16}$cm$^{-3}$) GaAs buffer layer was first grown by Molecular Beam Epitaxy on an n+-GaAs(001) substrate. The As(2*4)-reconstructed GaAs(001) surface was then exposed to 10 000 Langmuirs of HDT under UHV through a leak valve. After a 1 hour gentle annealing the sample at 373K in order to remove physisorbed molecules, the Co top contact (thickness 3nm) was deposited at a rate of 0.89 Å/min, either with sample maintained at room temperature (RT), or using a soft landing deposition process [42]. The Co Knudsen cell was distant from sample surface from 13 cm and operating during deposition with a total power of 9.8

W. For the soft-landing Co deposition, the SAM/GaAs(001) sample was cooled down at 25K and exposed to 900 Langmuir of Xe, forming a 100nm-thick protective Xe ice layer on the SAM before Co deposition. A 4nm-thick Au cap layer was finally grown before slowly warming sample up to RT with a ramp of 0.5K/min to desorb the Xe layer. All RT-deposited or soft landed top Au/Co top contacts were grown either full plate on the SAM/GaAs(001) surface for XPS characterization or patterned in circular dots (diameter 300µm) by using shadow mask deposition for BEEM measurements. Reference Au(4nm)/Co(3nm)/GaAs(001) Schottky contacts without SAM were also grown using the same protocol at RT or using soft-landing for comparison with samples integrating the SAM tunnel barrier.

Finally Au/Co/SAM/Fe(001)/MgO(001) magnetic tunnel junctions were patterned in the cross-bar geometry by using shadow mask deposition. 49 junctions could be patterned on a $1 \times 1 cm^2$ MgO(001) substrate while maintaining the complete deposition process under UHV. The geometry of the junction (length/width/thickness of the top and bottom electrodes) was designed to achieve tunnel junction resistance values significantly larger than the top and bottom metal electrode resistance. The MgO(001) substrate was first cleaned ex situ in an $O_3$ generator and degassed 1h under UHV at 860K. A monocrystalline Fe(001) electrode (thickness 30 nm) was then deposited at room-temperature on MgO(001) by Molecular Beam Epitaxy (MBE) and subsequently annealed at 860K to achieve a high-crystallinity and low-roughness Fe(001) surface as observed by low-energy electron diffraction and STM (see supplementary information figures S1 and S2). The HDT SAM grafting procedure of the Fe(001) surface was the same as on the GaAs(001) surface. A dense HDT SAM was obtained as checked by XPS analysis. The Au(4nm)/Co(15nm) top electrode was subsequently deposited by soft landing or at RT on the HDT SAM.

**Characterization.** X-ray photoemission spectroscopy (XPS) experiments were performed in an analysis chamber connected to the metal deposition and grafting chambers to investigate the top Co electrode growth and SAM molecular grafting quality. A dual non-monochromatized photon source with photon energies of 1253.6 eV (Mg Kα) or 1486.6 eV (Al Kα) was used while a hemispherical photoelectron analyzer (VSW HA100) working at an energy pass of 20eV provided an experimental energy resolution of 1.0eV. The binding energy scale was calibrated on the $Au4f_{7/2}$ peak measured on a reference clean Au surface and set at 84.0±0.1eV.

Ballistic Electron Emission Microscopy (BEEM) measurements were conducted to assess the spatial homogeneity of the HDT organic tunnel barrier after Co top contact deposition. The BEEM setup is based on a modified Omicron Scanning Tunneling Microscope (STM) operating at room temperature in the constant-current mode, with tunneling currents typically set to 10 nA.

Electrical transport measurements were carried out at 77K in a Janis cryostat under secondary vacuum of approximately $10^{-6}$ mbar using a Keithley 6482 picoammeter. Tungsten tips connected to triaxial cables were used to contact the top/bottom electrodes of the MTJs.

**ACKNOWLEDGMENTS**

The authors acknowledge the support of the French National Research Agency (ANR) (HITS project, ANR-20-CE09-0016). This work was also supported by France 2030 government investment plan managed by the French National Research Agency under grant reference PEPR SPIN – SPINMAT ANR-22-EXSP-0007. P.T. gratefully acknowledges Rennes Métropole for financial support of the BEEM setup development.